\documentclass[numberedappendix,appendixfloats]{emulateapj}

\shorttitle{}
\shortauthors{Fraschetti}

\usepackage{color}

\usepackage{amssymb,amsmath}
\usepackage{epstopdf}
\begin{document}

\title{Turbulent amplification of magnetic field driven by dynamo effect at rippled shocks}

\author{F. Fraschetti\altaffilmark{1}}
\affil{Departments of Planetary Sciences and Astronomy, University of Arizona, Tucson, AZ, 85721, USA}

\altaffiltext{1}{Associated Member to LUTh, Observatoire de Paris, CNRS-UMR8102 and Universit\'e Paris VII,
5 Place Jules Janssen, F-92195 Meudon C\'edex, France.}

\begin{abstract}  %600 characters spaces 

We derive analytically the vorticity generated downstream of a two-dimensional rippled hydromagnetic shock neglecting fluid viscosity and resistivity. The growth of the turbulent component of the downstream magnetic field is driven by the vortical eddies motion. We determine an analytic time-evolution of the magnetic field amplification at shocks, so far described only numerically, until saturation occurs due to seed-field reaction to field lines whirling. The explicit expression of the amplification growth rate and of the non-linear field back-reaction in terms of the parameters of shock and interstellar density fluctuations is derived from MHD jump conditions at rippled shocks. A magnetic field saturation up to the order of milligauss and a short-time variability in the $X$-ray observations of supernova remnants can be obtained by using reasonable parameters for the interstellar turbulence.

\end{abstract}


\section{Introduction} 

Compelling evidence has been cumulated that individual shell-type 
Supernova Remnant (SNR) shocks accelerate charged particles, 
i.e., electrons and probably ions, up to energies at least 
of the order of $10^{13} -10^{14}$ eV (see e.g. Cassiopeia A \citep{a01}, 
RX J1713.7-3946 \citep{a04}, Tycho's SNR \citep{a11}).
Charged particles are likely to be accelerated by two simultaneous mechanisms: 
the so-called Fermi first-order, i.e., repeated shock crossing of the particle \citep{a77,b78a,b78b,bo78,k77}, 
and drift along the shock surface \citep{j82,j87}.
From detection of non-thermal $X$-ray rims \citep{vl03,byuk04},
rapid time-scale variability of $X$-ray hot spots \citep{u07} 
and $\gamma$-ray emission in extended regions \citep{a11}, 
a magnetic field at the shock far exceeding the theoretically predicted shock-compressed field has been inferred. 
The $X$-ray rims could be due to damping of the magnetic field \citep{pyl05}  
rather than to synchrotron emission, although corresponding 
narrow filaments in the radio emission have not been observed \citep{r04} 
(see however high-resolution radio images in \citet{dcm09}).
Magnetic field amplification might be also relevant to {\it in situ} measurements 
of the plasma downstream of the solar-wind termination shock \citep{b07}, 
where fluctuations have been measured of the same order as the mean,
or to radio observations of Mpc scale shocks at the edge of galaxy clusters \citep{b12}.
Whether or not such a magnetic field amplification in SNR is to be associated with 
energetic particles at the shock is still subject of controversy.

So far, SNR observations could not rule out either of the two following mechanisms 
of field amplification: 
a) microscopic plasma instabilities generated by cosmic-rays current 
flowing upstream of the shock and therein exciting non-resonant magnetic modes \citep{b04}; 
b) macroscopic turbulent fluid motion downstream of the shock seeded
by inhomogeneities of the upstream medium triggering vortical eddies 
and tangling the magnetic field lines hence amplifying the turbulent component \citep{gj07}
(cosmic-rays pressure gradient has also been proposed as driver of the amplification in \citet{df86}).
The former mechanism was also discussed in the kinetic theory approach \citep{ab09}; however, 
numerical simulations (see, e.g., \citet{rs09}) could find only a moderate amplification ($B/B_0 \sim {\mathcal O}(10)$), 
and the unfolding of its non-linear extension and observational implications is currently active.

Collisionless shocks propagate in turbulent and inhomogeneous media undergoing 
rapid corrugation of their ideal planar surface.
Numerical simulations have shown that the picture of a planar shock is inappropriate 
to describe secular evolution of downstream medium. The unshocked medium might be 
strongly inhomogeneous at several scales and the cold interstellar clumps strongly deform the shock surface. 
The passage of an oblique non-relativistic shock through inhomogeneous medium has been known 
for longtime to generate vorticity in the downstream flow \citep{i64}; in a conducting fluid
the turbulent motion at scale ${\it l}$ with fluid velocity $v_{\it l}$ and local density $\rho$ 
leads to an exponentially amplified magnetic field $B^2 = 4\pi\rho v_{\it l}^2$ \citep{ll60}.
Such a dynamo action amounts to a systematic conversion of the fluid kinetic energy 
into magnetic energy at each scale separately, possibly until equipartition is reached \citep{k05}.
Recent numerical 2D-MHD simulations 
have shown that 
such an amplification can be very efficient \citep{gj07}. 
Further numerical studies using the thermal instability of the interstellar medium (ISM) turbulence,
i.e., condensation of the interstellar gas due to catastrophic radiative cooling \citep{f65},
confirmed the efficient magnetic field growth (\citep{i12} and references therein). 
Two-dimensional simulations of relativistic shocks \citep{m11} show that 
small-scale dynamo can operate also downstream of the shocks of Gamma-Ray Bursts outflows, 
suggesting that the dynamo action downstream of shocks 
might shed light on the energy equipartition at magnetized shocks. 

In this paper we provide an analytic derivation of the vorticity generated by clumpy 
unshocked medium, hence of the magnetic field amplification, downstream of a non-relativistic 
rippled collisionless shock. 
We apply the Rankine-Hugoniot jump conditions locally 
downstream of an MHD shock to compute the vorticity generated downstream. 
For the sake of simplicity, a two-dimensional shock is considered, i.e., 
observables depend only upon two space coordinates.
The downstream vorticity depends on the magnitude of the tangential component of the velocity 
(shear), although a different interpretation is given here in terms of density gradient at clumps boudary, and
the curvature of the shock surface as previously found for purely hydrodynamic shocks  \citep{t52,k97}. 
We also compute the back-reaction to vortical motion of the seed magnetic field advected downstream, 
so far accounted for only numerically \citep[e.g.][]{gj07,i12}. 
Using the small-scale dynamo theory we determine the time-evolution 
of the turbulent magnetic field in the downstream fluid until saturation epoch.

The encounter of a shock surface with a density clump, also called 
Richtmyer-Meshkov (RM) instability \citep{b02}, has been extensively investigated in 
plasma laboratory experiments (see \citet{dr10} and references therein). 
Numerical simulations of magnetic shocks proved that RM instability drives 
transient events in several regions of the Earth magnetosphere \citep{wr99}. 
Recent plasma laboratory experiments \citep{k11} made use of laser to test 
the magnetic field amplification by density inhomogeneities at shocks 
of supernova remnant (see \citet{s12} and references therein).

The small-scale fluid vortices 
close behind the shock grow on time-scale smaller than the particle acceleration time-scale, 
which depends on the seed magnetic field orientation, the isotropy of the turbulent component of magnetic field 
and the dependence of the spatial diffusion coefficients, parallel and perpendicular, on the particle energy.
Therefore, the vortical field growth is unaffected by the presence of cosmic-rays at the shock.
Magnetic field may also be enhanced by field line stretching 
due to Rayleigh-Taylor (RT) instability \citep{jns95} at the interface between the ejecta 
and the interstellar medium, i.e., far downstream of the shock.
In contrast with the vortical turbulence, late-time RT turbulence might be affected 
by the highest energy particle gyrating in the downstream fluid
far from the shock \citep{ftbd10}. However, RT structures are unlikely to reach out the blast wave 
(\citep{ftbd10} and references therein) and therefore to interact with vortical turbulence.
Thus the dynamo amplification local behind the shock can be temporally and spatially disentangled 
from the field line stretching due to RT instability.

This paper is organized as follows: in Sect. \ref{sect_rippled} we introduce the constitutive equations and describe the features of a  rippled shock. In Sect. \ref{sect_vorticity} we define the vorticity in the local rotated frame, compute the vorticity downstream of the shock and interpret the result in terms of vorticity growth and field back-reaction. In Sect. \ref{sect_amplif_field} we apply the small-scale dynamo theory to determine the time-evolution of the turbulent magnetic field. In Sect. \ref{sect_discuss} we identify the  dependence of field growth and non-linear field back-reaction on the physical shock parameters and discuss the implications for recent $X$ and $\gamma$-ray observations of non-relativistic SNR shocks. In Sect. \ref{sect_concl} we summarize our findings.

\section{Rippled shock}\label{sect_rippled}

We consider the propagation of a 2D non-relativistic shock front in an inhomogeneous medium. The time evolution of the two driving independent observables in ideal MHD approximation, i.e., the fluid velocity ${\bf v} = {\bf v} (x,y)$ and the magnetic field ${\bf B} = {\bf B} (x,y)$, is given, with no viscosity or heat conduction and for infinitely conductive fluid, by
\begin{equation}
{\partial_t {\bf v}}+({\bf v}\cdot\nabla){\bf v} + \frac{\nabla P}{\rho} + \frac{1}{4\pi\rho} \left[ {\bf B}\times(\nabla \times {\bf B}) \right]  = 0
\label{Euler}
\end{equation}
\begin{equation}
{\partial_t {\bf B}}=\nabla\times({\bf v}\times{\bf B}) 
\label{induction}
\end{equation} 
where $\rho$, $P$, are respectively density and hydrodynamic pressure of the fluid (here $\partial_{t} = \partial / \partial t$). Equation (\ref{induction}) does not include any field-generating term, such as Biermann battery \citep{k05},  as the fluid on both sides of the shock is embedded in a pre-existing magnetic field. Note that the current density carried by cosmic-rays is here neglected: we aim to identify the growth of the magnetic energy as generated by the vortical motion of the background fluid only. 
Thermal dissipation reduces the energy deposited in the magnetic turbulence and will be considered in a forthcoming publication. 

\begin{figure}
\includegraphics[width=8cm,height=6cm]{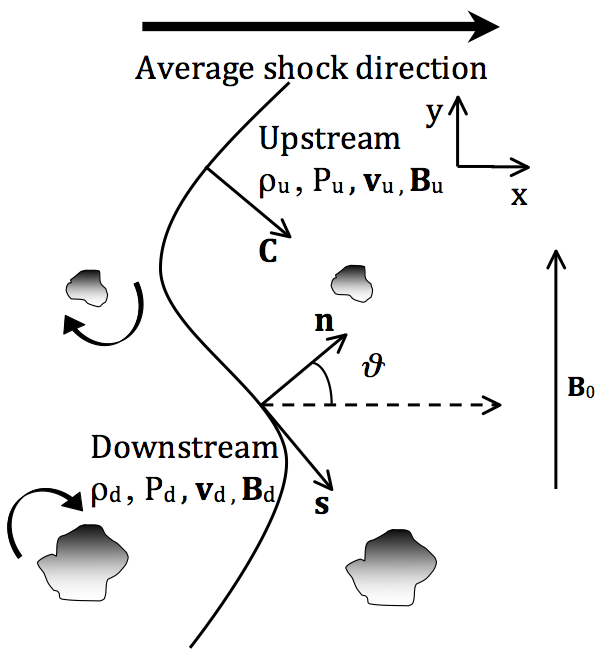}
\caption{Encounter of a shock surface with density enhancement regions: forward and lagging behind regions are formed that generate vorticity in the downstream fluid. } 
\label{fig}
\end{figure}

The shock is a dynamic surface due to the interaction with the upstream clumps.
The kinematics of a 2D hydrodynamic rippled shock propagating in inhomogeneous medium \citep{rp93}  comprises of a sequence of unstable configurations of the shock surface. The conservation of energy requires that sections of the corrugated shock surface being ahead or lagging behind readjust to smooth-out growing corrugations. The time-evolution of the shock surface is accounted for by the rate change along the moving surface \citep{p01} of $\vartheta(t,x,y)$, i.e., the local angle between the average direction of the shock motion and the local normal to the shock surface (see Fig.\ref{fig}). We assume here that such a self-deformation process of the shock surface occurs on a time-scale much greater than the turn-over time of the smallest eddies in the downstream flow of the shock. Thus the shock profile is ``frozen'' during the exponentially fast amplification which proceeds as the field is advected downstream.

\section{Vorticity downstream of MHD shock}\label{sect_vorticity} 

In the inviscid approximation used here the conservation of the vorticity flux applies: the vorticity shock-generated is transported along the flow ``frozen'' into the fluid, as a consequence of Helmholtz-Kelvin theorem.
The medium upstream of the shock has zero vorticity. The vorticity is calculated downstream at a distance from the shock large enough that the shock is infinitely thin, i.e., the thickness of the shock is much smaller than the local curvature radius at every point of the shock surface. 

At a rippled shock the MHD Rankine-Hugoniot jump conditions cannot be applied globally as the directions normal and tangential vary along the shock surface. In a 2D shock propagating at average in the direction $x$ (Fig.\ref{fig}), from the velocity field of the flow ${\bf v} = (v_x, v_y, 0)$, the vorticity is given by ${\boldsymbol{\omega}} = \nabla \times {\bf v} = (0, 0, \omega_z)$ and the component in the direction $z$, outgoing from the paper, {\bf by} $\omega_z = \partial_x v_y - \partial_y v_x$ (all quantities are independent on $z$ and we used $\partial_{x_i} = \partial/\partial_{x_i}$). We use a local natural coordinate system $(\hat n,\hat s)$, where $\hat n = ({\rm cos} \vartheta(t, s), {\rm sin} \vartheta (t, s))$ is the coordinate along the normal to the shock surface, $\hat s = ({\rm sin} \vartheta (t, s), {\rm -cos} \vartheta (t, s))$ is the coordinate parallel to the shock surface (Fig.\ref{fig}). Note that $\vartheta =\vartheta (t, s)$ 
varies only along the shock surface and not along the orthogonal direction $n$, thus is independent on $n$ (also, the time-evolution of $\vartheta$ in \citet{rp93}, Eqs. [2.21]-[2.23], does not depend on $n$).  
In the local frame $(\hat n,\hat s)$ the z-component of the vorticity becomes 
\begin{equation}
\omega_z = \partial_s v_n - \partial_n v_s + v_s \partial_s \vartheta ,
\label{omegans} 
\end{equation}
where $v_n = v_x {\rm cos} \vartheta + v_y {\rm sin} \vartheta $ and $v_s = v_x {\rm sin} \vartheta - v_y {\rm cos} \vartheta$ are respectively the local component of the velocity flow normal and tangential to the shock (see Fig.\ref{fig}). The local derivatives here can be expressed in terms of derivatives in directions $(\hat x, \hat y)$ as $\partial_n = \hat n \cdot \nabla = {\rm cos} \vartheta~\partial_x + {\rm sin} \vartheta~ \partial_y $ and $\partial_s = \hat s \cdot \nabla = {\rm sin} \vartheta~\partial_x - {\rm cos} \vartheta~ \partial_y $. The vorticity generated downstream of the shock in the presence of a magnetic field ${\bf B } = (B_n, B_s, 0)$, for $\omega_z = 0$ upstream (detailed computation is in the Appendix), results in   
\begin{align}
\delta \omega_z & = -\frac{1}{\rho C_r} \left\{ \frac{r-1}{r} [C_r]_u \partial_s (\rho C_r)   \right.\nonumber\\
& \quad \left. -  \frac{\partial_s \delta B^2}{ 8\pi}  +    \frac{B_n \delta B_s}{4\pi}\partial_s \vartheta + \frac{B_n}{4\pi} \delta[\partial_s B_n - \partial_n B_s ]  \right\} ,
\label{deltaomega2}
\end{align}
where $C_r = C - v_n$ is shock speed in the fluid frame ($C$ is the shock speed in the normal direction $n$), $r=\rho_d / \rho_u$ is the compression ratio at the shock and $\delta f = [f]_d - [f]_u$ indicates the jump of $f$ from upstream ({\it u}) to downstream ({\it d}).

In the present paper, we deal with the hypothesis that the turbulent magnetic component is only generated in the downstream flow through dynamo mechanism, i.e., the turbulence is not required on both sides of the shock as in the diffusive shock acceleration. The contribution to strong dynamo amplification in the downstream flow from far-upstream fluctuations of magnetic and flow velocity fields on several scales will be included in a forthcoming work. If the magnetic field is quasi-perpendicular to the average direction of shock motion, charged particles can be still efficiently accelerated to high-energy by the motional electric field through drifting along the shock, until they are advected in the downstream flow or escape upstream, provided they are not scattered back to the shock to further acceleration. 
In an oblique or quasi-parallel configuration, if the field is amplified only in the downstream flow as described here, other upstream turbulence or  pre-existing magnetic instabilities are needed to scatter energetic particles back across the shock, increase the residence time in the upstream region and release larger energy accelerated particles. Since no upstream magnetic turbulence is included here, at the present stage this model reconciles dynamo amplification with particle acceleration occurring in the same region of the shock only for quasi-perpendicular magnetic field. Turbulence of upstream medium allows an extension to oblique and quasi-parallel cases. 

We consider a seed-magnetic field upstream uniform and normal to the average direction of motion (${\bf B}_0 = (0, B_0^y, 0)$, or $B_n=~B_0 {\rm sin} \vartheta$ and $B_s=-B_0 {\rm cos} \vartheta$, see Fig.\ref{fig}). For the first term in the second line of Eq. (\ref{deltaomega2}), we can write $\partial_s \delta B^2 = \partial_s \delta B_s^2 \sim -2 B_0^2 \delta ({\rm sin}2\vartheta \partial_s \vartheta) =0$, as $ \delta ({\rm sin} \vartheta) = \delta (\partial_s \vartheta) = 0$. The last term in Eq.(\ref{deltaomega2}) vanishes: $\delta[B_0 {\rm cos} \vartheta (\partial_s \vartheta)] = 0$ and $\partial_n B_s =0$. 

Assuming that amplification is efficient at the smallest scales (see Sect. \ref{sect_amplif_field}), $B_n$ and $\delta B_s$ are drowned out by the turbulent components: $\delta B_s \sim -rB_0 -B_s^{turb} +B_0 \sim -B_s^{turb}$ and $B_n \sim B_n^{turb}$ for a perpendicular field (see also last paragraph in the present Section). Therefore the factors $\delta B_s $ and $B_n$ in Eq.(\ref{deltaomega2}) include both the impulsive shock compression and the turbulence amplification. We can conclude that the vorticity produced downstream of a 2D shock propagating in an inhomogeneous medium with a uniform perpendicular upstream magnetic field (same as for parallel shock as shown later) can be recast, neglecting obliqueness, in a simple form:
\begin{equation}
|\delta \omega_z|  =  \frac{r-1}{r} \left[ \left(\frac{C_r}{\rho}\right)_u \partial_s \rho + \partial_s C_r \right]  -  \frac{B_n \delta B_s}{4\pi \rho C_r} \partial_s \vartheta ,
\label{dom_perp}
\end{equation}
where $\delta B_s$ is the jump across the shock of the magnetic field in the direction locally tangential to the shock surface including the Rankine-Hugoniot compressed seed field and the turbulently amplified field and $B_n$ is the component in the direction locally normal to the shock surface including the unchanged Rankine-Hugoniot and the turbulent component. 

The factor $(r-1)/r$ shows the dependence of $\delta \omega_z$ on the shock compression. The first term in Eq.(\ref{dom_perp}) contains the density gradient ($\partial_s \rho$), hence having the role of the baroclinic generation of vorticity, i.e., the term $\nabla P \times \nabla \rho$ appearing in the equation for $\boldsymbol{\omega}$ time-evolution: when an upstream clump (clump size is assumed greater than shock thickness) crosses the shock the clump is locally heated with a maximum compression at the front of the cloud and smaller along the side of the cloud, whereas the density gradient is directed across the fluid-clump interface \citep{kmc94}: thus, the gradients of density and pressure are no longer parallel and vorticity is generated in the transition layer ($\nabla\rho \neq 0$). The thermal instability model for ISM predicts a broad range \citep{f65,bm90} for the transition layer between the cloud and the ISM shocked gas as a function of the thermal conduction (see Sect. \ref{sect_discuss}).  
The second, or corrugation, term in Eq.(\ref{dom_perp}) results from the finite curvature radius of the shock, i.e., for a planar shock $\partial_s C_r = 0$ at every point of the shock surface. As shown in the following Section, the purely hydrodynamic terms (baroclinic and corrugation) drive the small-scale magnetic field growth.

The back-reaction of the small-scale turbulent field is represented by the last term: vortical eddies shear and whirl field lines around enhancing the turbulent field as long as the entailed magnetic tension grows to strength large enough to halt such a growth. Similar dependence of the field back-reaction on the turbulent-field Alfv\'en speed can also be reasonably derived by dimensional arguments \citep{k05} but it follows here from the application of local Rankine-Hugoniot jump conditions at rippled shocks. Note that in the absence of shock ripples, i.e., for infinitely large local curvature radius, the term $ \partial_s \vartheta \rightarrow 0$ and back-reaction of the amplified field on the vorticity becomes negligible. Also, the vorticity downstream of a shock wave in Eq.(\ref{dom_perp}) holds regardless the speed of the shock wave (Alfv\'en, fast or slow magnetosonic  speed). We recall that in contrast with 3D turbulence, the shear of the fluid velocity along the vorticity, i.e.,  $({\boldsymbol{\omega}} \cdot \nabla) {\bf v}$, vanishes in a 2D flow and the vorticity generation in a 2D fluid might be under-estimated.

Note that if the magnetic field upstream is uniform and parallel to the average direction of motion (${\bf B}_0 = (B_0^x, 0, 0)$, or $B_n=B_0 {\rm cos} \vartheta$ and $B_s=B_0 {\rm sin} \vartheta$), the magnetic term in Eq.(\ref{dom_perp}) is unchanged. Therefore, assuming efficient turbulent field amplification, a difference in the growth of $|\omega|$, i.e., in the saturated turbulent field, is not expected between the perpendicular and parallel field upstream cases, confirming previous numerical findings \citep{i12} (see also Sect. \ref{sect_amplif_field}). The reason is that the turbulent field isotropically becomes much greater than $B_0$, regardless the orientation of the seed-field upstream.

%==================

\section{Turbulent field amplification} \label{sect_amplif_field}

As a result of dynamo action \citep{k05}, the growth of a turbulent magnetic field is known to be governed by the fluid vorticity at each scale. The unperturbed field is initially too weak to affect the fluid velocity field and the turbulent field grows exponentially fast, i.e., on the time scale of the smallest eddies turnover time \citep{bk05}, until the magnetic energy produces non-negligible effects on the velocity field and then saturates. 

The time-evolution of the total magnetic field (Eq.\ref{induction}), i.e., both ordered and turbulent components, is determined here by using the small-scale dynamo theory \citep{k05}. The vortical swirling exponentially amplifies also the mean field. However, since the amplification time-scale is of the order of the smallest eddies turnover time, the saturation occurs much faster at small-scale. 
No statistical assumption is made on the magnetic field, in contrast with the  usual numerical approach. This is a relevant simplification as the spatial diffusion of charged particles in a turbulence depends on the magnetic power spectrum (see, e.g., \citet{fj11}). 

The small-scale dynamo theory predicts that the turbulent field obeys an unbounded exponential amplification at a rate $\beta$ \citep{k05,ka12}: $d\varepsilon /dt = 2\beta \varepsilon$, where $\varepsilon = B^2 / 8\pi \rho$ is the total magnetic energy per unit of mass, including seed and turbulent fields. As shown in \citet{k05}, the isotropy and homogeneity of the velocity correlation entails the following simple relation between the amplification rate of $\varepsilon$ and the vorticity generated downstream of the shock: $\beta \simeq (\pi/3) \delta \omega_z$. However, such a version of the dynamo theory cannot resolve the dilemma posed by the unlimited magnetic growth and several non-linear versions have been elaborated \citep{k05}. 

We present in this paper the back-reaction of the field to the vorticity growth and the saturation of $B$ at realistic rippled shocks. We argue that Eq.(\ref{dom_perp}) for $|\delta \omega_z|$ allows to explore the regime of self-controlled growth wherein the turbulent field becomes so strong to affect the velocity field. The time-evolution of the turbulence amplification until the saturation epoch can be  determined: if we recast Eq.(\ref{dom_perp}) as $ |\delta \omega_z| = (3/\pi) (\tau^{-1} - \alpha \varepsilon)$, then $\varepsilon$ satisfies 
\begin{equation}
\frac{d \varepsilon}{dt} = 2 (\tau^{-1} - \alpha\varepsilon)\varepsilon
\label{epsilonEq}
\end{equation}
where $\tau^{-1} = \frac{\pi}{3} \frac{r-1}{r} \left[ (C_r/\rho)_u \partial_s \rho + \partial_s C_r \right] $ is the local growth rate of $\varepsilon$ and $\alpha =  (2 \pi/3) \partial_s \vartheta /C_r $ is the local back-reaction; the initial condition for Eq. (\ref{epsilonEq}) is $\varepsilon (0) = \varepsilon_0 = v_A^2 / 2= B_0^2/8\pi\rho $. In Eq.(\ref{epsilonEq}) we have assumed that the turbulence dominates over $B_0$, i.e., $\delta B_s/\sqrt{8\pi \rho} \sim \sqrt{\varepsilon}$ and $ B_n/\sqrt{8\pi \rho} \sim \sqrt{\varepsilon}$: the turbulence grows isotropically downstream at the shock curvature scale as a consequence of the isotropy of the flow velocity field \citep{k05}. We note that the ISM clump size is much larger than the shock thickness and the shock crossing is not impulsive on the vorticity generation time-scale; thus, $\varepsilon$ in the bracket of Eq.(\ref{epsilonEq}) is continuously generated during the shock crossing and is not constant in time. 

Neglecting the time dependence of $\tau$ (the magnetic modes grow slowly for initially weak field \citep{k05}), the solution is readily found: 
\begin{equation}
\frac{\varepsilon}{ \varepsilon_0} (t) = \left(\frac{B}{B_0}\right)^2 (t) = \frac{e^{2t/\tau}}{1-\alpha\tau(1-e^{2t/\tau})v_A^2 /2},
\label{exact}
\end{equation}
for a uniform average interstellar matter density. Equation (\ref{exact}) provides the first analytic time-evolution of a self-controlled amplification of magnetic energy through fluid vortical motion generated downstream of a 2D rippled shock. 

For $t \ll \tau$ the term proportional to $v_A^2$ in the denominator is negligible and $B$ grows exponentially ($B/B_0 \sim e^{t/\tau}$). For times $t \gg \tau$ the term proportional to $v_A^2$ dominates over the first term and the saturation is attained: 
\begin{equation}
B/B_0  \simeq \sqrt{2/(\alpha \tau v_A^2)} ,
\label{saturation }
\end{equation}
corresponding to an increase $B/B_0 \sim M_A$, where $M_A = C_r \sqrt{4\pi \rho}/B_0$ is the seed-field Alfv\'en Mach number (see Sect. \ref{sect_discuss}). Note that this result applies only if the assumption that the turbulence $B^{turb}$ downstream dominates the compressed seed field $rB_0$. For interplanetary shocks ($M_A \lesssim 10$) the measured downstream turbulence is rarely amplified to values much greater than the mean field \citep{b07}, and the term $r B_0$ is no longer negligible with respect to $B^{turb}$ and a different analysis is needed. 

\section{Discussion}\label{sect_discuss}

Whether the magnetic field is amplified by waves excitation in the shock precursor or downstream through the coupling of the eddies motion with the magnetic field lines, or a synergy of the two mechanisms, or other processes, is not observationally settled yet. Only the former has been largely explored in recent years. This paper presents for the first time an analytic approach to the latter mechanism.

In what follows we estimate $\tau$ and $\alpha$ showing that Eq.(\ref{exact}) applies to the time-evolution of magnetic energy in young and middle-aged SNRs. We do not aim here at predicting a typical length-scale of the $X$-ray rims due to the large variety of SNRs taxonomy, but rather we show that reasonable values of the clump/fluid transition layer, shock curvature radius and shock speed are capable to produce the inferred efficient magnetic field amplification.

\begin{figure}
\includegraphics[width=9cm,height=5.5cm]{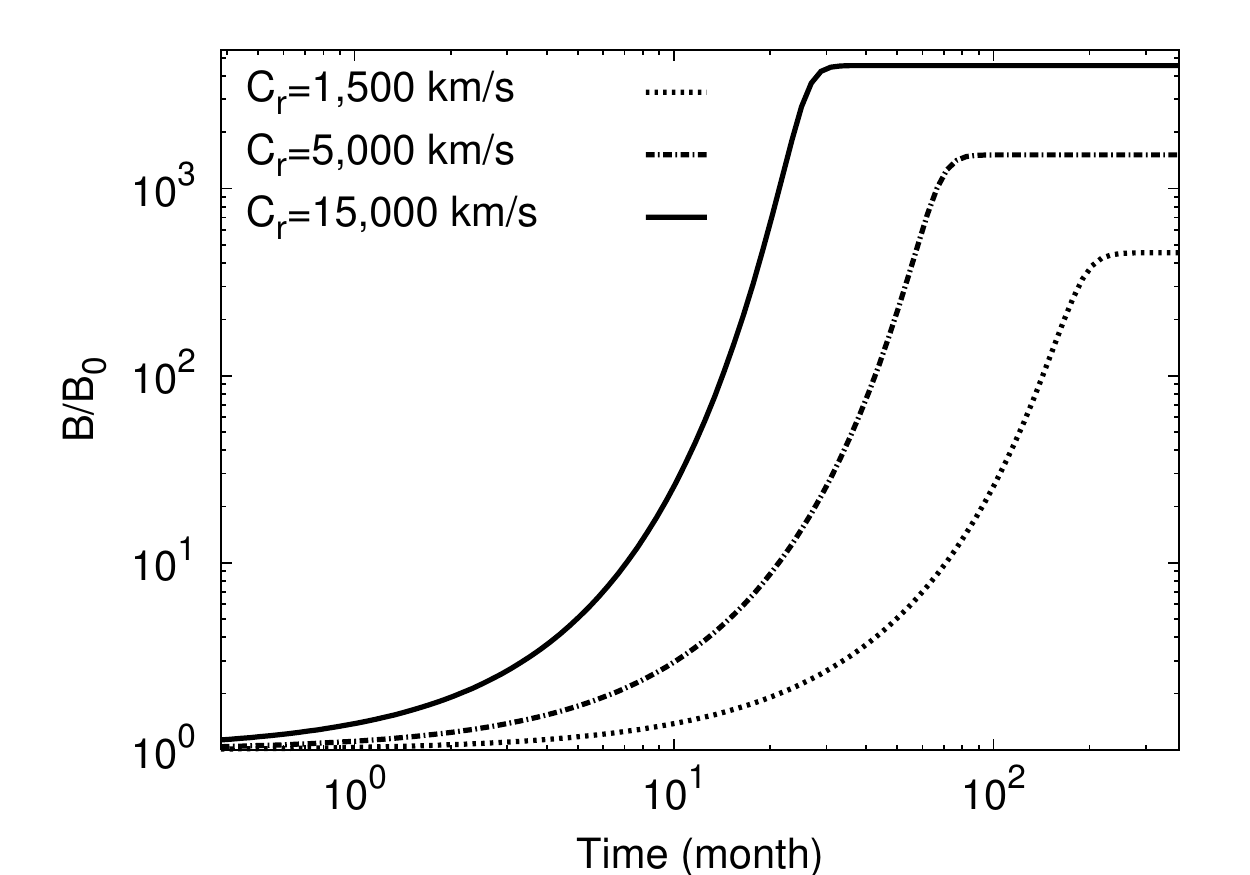}
\caption{Saturation of the total magnetic field for various shock speed $C_r$ is shown: $C_r = 1,500$ km$/$s ($M_A =50$), $C_r = 5,000$ km$/$s ($M_A \sim 170$), $C_r = 15,000$ km$/$s ($M_A =500$), assuming $R_c = 10^{17}$ cm and $\ell_F = 10^{16}$ cm, that results in $\tau \lesssim \ell_F/C_r \sim 3$ years for $C_r = 5,000$ km$/$s ($\vartheta = 0.1$ rad, $r=4$ and $v_A = 10^{-4} c$).}
\label{Cr}
\end{figure}

\begin{figure}
\includegraphics[width=9cm,height=6cm]{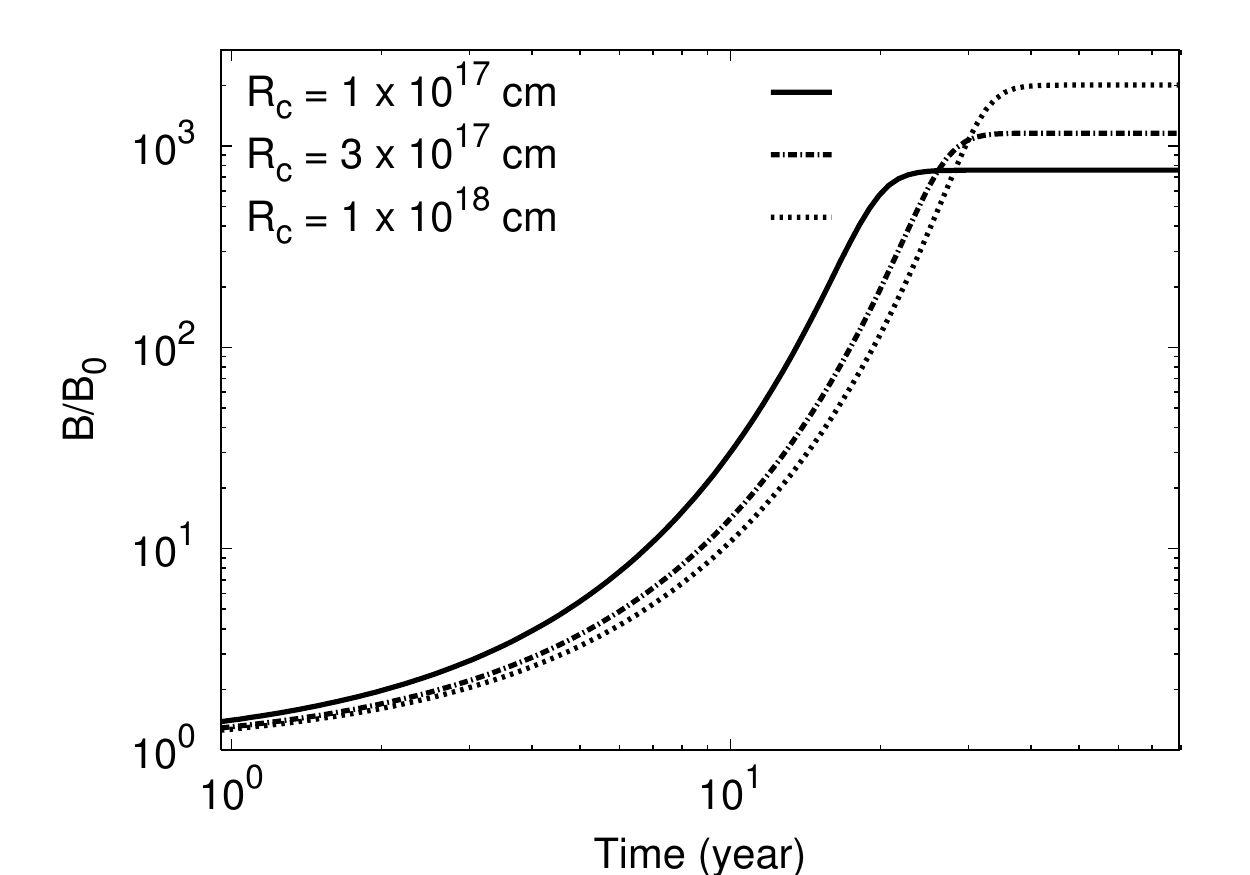}
\caption{Saturation of the total magnetic field for various $R_c$ is shown, for $C_r = 4,800$ km$/$s ($M_A \sim 160$), and using $\ell_F = 5 \times 10^{16}$ cm ($\vartheta = 0.1$ rad, $r=4$ and $v_A = 10^{-4} c$).}
\label{Rc}
\end{figure}  
  
\begin{figure}
\includegraphics[width=9cm,height=7cm]{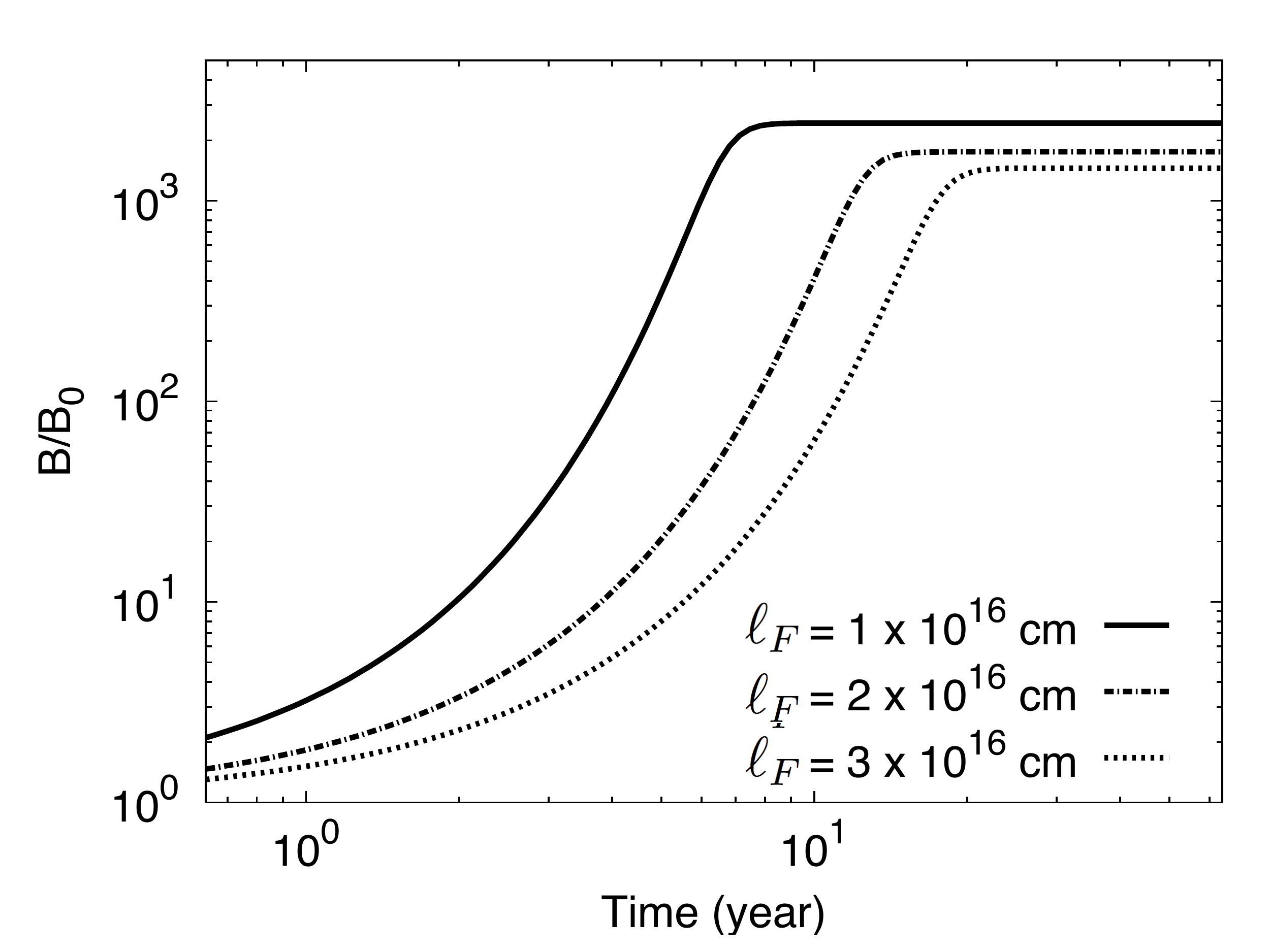}
\caption{Saturation of the total magnetic field for various values of $\ell_F$ for $C_r = 4,800$ km$/$s ($M_A =160$) and $R_c = 3 \times 10^{17}$ cm ($\vartheta = 0.1$ rad, $r =4$ and $v_A = 10^{-4} c$).}
\label{LF}
\end{figure}

\subsection{{\it Field growth}}

From the discussion below Eq.(\ref{dom_perp}), the magnitude of the vorticity downstream $|\delta \omega_z|$ is expected to depend on the thickness of the layer of the interface clump/fluid and not on the size of the clump. Thermal instability model of two-phase fluid suggests that such a thickness is given by the Field length $\ell_F$ \citep{f65,iik06}. However, $\ell_F$ depends on the heat conduction \citep{bm90} and is thus uncertain as the thermal conductivity might be sensitive to the magnetic field. The previous history of formation and evolution of the supernova progenitor might also affect the properties of the surrounding medium, i.e., uniform or stellar wind  density profile.

The growth time-scale can be recast as $\tau^{-1} = \frac{\pi}{3} \frac{r-1}{r} \left[ (C_r/\rho)_u \partial_s \rho + \partial_s C_r \right] \sim  \frac{r-1}{r} C_r (R_c^{-1} +\ell_F^{-1})$. Neither of the terms $R_c^{-1}$, $\ell_F^{-1}$ can be neglected because no particular assumption has been made on $R_c$, expected indeed to be comparable to the size of the clump that corrugated the shock, with respect  to $\ell_F$. From the previous argument it follows:
\begin{equation}
\tau \sim \frac{r}{r-1} \frac{1}{C_r} \frac{R_c \ell_F}{R_c +\ell_F} .
\label{tauEq}
\end{equation}
The growth rate $\tau^{-1}$ of the turbulent field, and of $\delta \omega_z$, increases with shock speed and it depends only on hydrodynamic quantities, except a possible unknown dependence on $B$ in $\ell_F$ here disregarded. If $\ell_F \ll R_c$, from Eq.(\ref{tauEq}) it holds $\tau \sim \ell_F/C_r$ (see also Fig.\ref{LF}). Thus, the field amplification saturates faster in a region of smaller $\ell_F$ providing a constraint testable by multiwavelength observations. This justifies the description in Sect.\ref{sect_vorticity} that the vorticity, and the strong turbulent field, are generated in the transition layer separating the cold clump from warm ISM and does not depend on the clump size.

Note that $\tau$ is smaller than the travel-time of the shock across the interior of the ISM clump, i.e., cloud-crushing time $t_{cc}$, resulting from the ram-pressure equilibrium at the clump boundary \citep{kmc94}: $t_{cc} = \sqrt{\rho_c/\rho_0} L/C_r$, where $\rho_c$ and $\rho_0$ are respectively mass-density in the clump and in the surrounding fluid. Structures of scale $L$ stable under external heating or radiative cooling condense if $L \gg \ell_F$; thus $t_{cc} = \sqrt{\rho_c/\rho_0} L /C_r \gg \sqrt{\rho_c/\rho_0} \ell_F /C_r \simeq \sqrt{\rho_c/\rho_0} \tau > \tau$. Thus, for any $\rho_c/\rho_0$ the ISM clump survives across the shock provided that the density gradient at the clump boundary is steep enough.

\subsection{{\it Non-linear back-reaction}}

As the magnetic field strengthens, it reacts to field lines whirling halting the turbulence growth. In more general terms, as the field increases by dynamo effect it also releases its tension by unwinding at a rate of order of Alfv\'en speed:  the backreaction grows with the turbulent field Alfv\'en speed \citep{k05}. The local back-reaction of the field $\alpha \sim \partial_s \vartheta /C_r $ can be estimated by 
\begin{equation}
\alpha \sim \vartheta/(R_c C_r) .
\end{equation}
Clearly $\alpha$ is enhanced by small curvature radius ($\alpha \sim 1/R_c$) as the more corrugated is the shock, the shorter the eddies turnover time, the more efficiently the turbulent field is created downstream and back-reacts to fluid motion.  From Eq.(\ref{epsilonEq}), it is clear that as $\varepsilon$ increases, the back-reaction term grows faster than the driving term, eventually becoming dominant: the strongest field has the smallest growth (cfr. the backreaction dependence on $B$ in the non-linear Landau damping).

\subsection{{\it Secular evolution of the turbulent field}}

Fig.\ref{Cr} depicts the growth of the turbulent field for various shock speeds, assumed constant in time \footnote{On the year time-scale the speed $C_r$ is approximately constant: in the ejecta-dominated phase ($t < 1,000$ yr) of a self-similar SNR contact discontinuity with radius $R \sim t^\lambda$, the speed $C_r \sim t^{\lambda-1}$ changes by a factor $\lesssim 5-6$.}: given an ISM field of the order of $B_0 \sim 3 \mu $G, the turbulent field saturates at $B \sim 1.2-3.$ mG for $C_r = 1,500 - 5,000$ km$/$s on the year time-scale. Such a rapid growth of magnetic energy is compatible with $X$-ray observations of SNRs $RX J1713.7-3946$ ($C_r < 4,500$ km$/$s \citep{u07}) and Cas A \citep{pf09} brightness variations detected on year time-scale in small-scale hot spots structures, attributed to synchrotron electron cooling. Using $R_c = 10^{17}$ cm and $\ell_F = 10^{16}$ cm, we find an amplification to $B \sim 3. $ mG within $3$ years. Such a value of $\ell_F$ is to be compared with the spatial scale of the {\it Chandra} $RX J1713.7-3946$ bright spots, estimated as $\lesssim 0,03$ pc. 
Similar length ($\sim 10^{14} - 10^{16}$ cm) and time ($\sim 1$ yr) scales are found in simulations of the effects of magnetic field turbulence on the observed synchrotron emission images and spectra in SNRs \citep{bue08}. Thus, the magnetic energy increase and the $X$-ray variability might have a time-scale ($1$ yr) much lower than the SNR hydrodynamic time-scale and might occur in middle-aged SNRs, not necessarily young SNRs ($RX J1713.7-3946$ age is estimated as $1,600$ yr \citep{sg02}). The high shock speed $C_r \sim 15,000$ km$/$s in Fig.\ref{Cr} is comparable to observations of the youngest SNR in our galaxy, i.e., $100$ years old G$1.9+0.3$ \citep{rbghhp08}. Thus, a rapid field saturation even up to $B \sim 10$ mG is predicted at SNR shocks within a few months.
 
The computation above indicates a linear increase of the saturation value with the shock speed (see Fig.\ref{Cr}), or $M_A$. From Sect. \ref{sect_amplif_field} the saturation can be written as 
\begin{equation}
\hspace{-0.46cm} \frac{B}{B_0}  \simeq \sqrt{\frac{2}{\alpha \tau v_A^2}} \sim \sqrt{2\frac{C_r^2}{v_A^2} \frac{R_c + \ell_F}{R_c \ell_F \partial_s \vartheta} } \sim M_A \sqrt{2\frac{R_c + \ell_F}{\ell_F \vartheta}} .
\label{scaling}
\end{equation}
The equipartition between the magnetic pressure of the turbulent component downstream of the shock, the thermal pressure and the fluid ram pressure ($B^2/8\pi \sim \rho C_r^2 $) implies: $(B/B_0)^2 \sim (8\pi/B_0^2) (\rho C_r^2) \sim 2 M_A^2$. Thus the scaling $B/B_0 \sim M_A$ indicates that the amplified field pressure might locally become comparable with the ram pressure (and the gas pressure), according to the factor $(R_c + \ell_F)/(\ell_F \vartheta)$: the non-linear back-reaction, found using jump conditions at rippled shocks in this paper, is no longer negligible (cfr. \citet{b04}) and the rates of growth and unwinding are equal.  
 
Fig.\ref{Rc} shows the turbulent field growth for various values of $R_c$, with a shock speed $C_r = 4,800$ km$/$s (close to the highest measured for the SNR1006 shock \citep{k12}). From {\it HST} observations of SN1006 \citep{rksbgs07}, $R_c$ is inferred to be less than $1/10$ of the forward shock radius, estimated as $R_s \sim 9.1 $ pc $= 2.8 \times 10^{19}$ cm. However, the overall smoothness of SN1006 suggests that the ripples scale might be lower ($R_c \sim 10^{16}$ cm) in more clumpy-structured SNRs. This justifies the smaller value of $R_c$ used in Fig.\ref{Cr}. Moreover, the absence in SN 1006 of hot spots with rapid time-variability, as detected in synchrotron cooling regions of $RX J1713.7-3946$ \citep{u07}, might result from the spatial location at high Galactic latitude where small-scale clumps are not present (see also \citet{k10}). 
Thus, a saturation $B/B_0 \sim 500-800 $ for $R_c = 10^{17}$ cm is compatible with the optical constraints on SN1006 \citep{rksbgs07}, indicating that a region with strong amplified field variable on $10$ yrs time-scale could still be observed in future high-spatial resolution efforts before the advected amplified field is drowned by projection effects. 

Despite several uncertainties, a numerical estimated range for $\ell_F$ in the ISM is given by $[10^{16} - 10^{18}]$ cm (see Eq.s [5.1], [2.7] in \citet{bm90}). The field growth is clearly more efficient at the small scale $\ell_F$, as also shown in Fig.\ref{LF}. In Fig.\ref{Rc} an increase in $R_c$ reduces the field back-reaction ($\alpha \sim 1/R_c$) resulting in a larger saturation value. In contrast (Fig.\ref{LF}) an increase in $\ell_F$ ($\ell_F \ll R_c$), reduces the growth rate ($\tau \sim \ell_F/C_r$) and therefore $B/B_0$.

ISM density turbulence has been known for long-time through radio scintillation to obey the Kolmogorov scaling over several decades in length-scale \citep{lj76,ars95} from the outer scale, or injection scale, $L_c \sim 3-50$ pc \citep{n12}. 
If the turbulent fluid velocity downstream of the shock follows a 1D Kolmogorov power spectrum at a certain scale $k^{-1}$ as numerical simulations seem to suggest \citep{i12} ($P_v(k) \sim k^{-5/3}$), the vorticity spectrum at scale $k$ is given by $P_{\omega} (k) = k^2 P_v(k) \sim k^{1/3}$: in the inertial range the vortical energy at small-scale is greater than at large scale. Also, since the field at small-scale grows and saturates faster, at that scale the magnetic energy coincides with the turbulent energy \citep{k05}. We indicate here such a scale to be $\lesssim 0.01$ pc. However, as a result of the uncertainties on the Field length $\ell_F$ \citep{f65}, within the scenario presented here a unique length-scale of magnetic growth cannot be identified. A range of scales consistent with the ISM fluctuations is shown here to account for optical and $X$-ray observations. The confirmation will be sought through kinetic analysis or 3D numerical simulations in forthcoming publications. 

An attractive feature of the microinstabilities mechanism in \citet{b04} is that magnetic field growth is associated with energetic particles at the shock and the field saturation is driven by the cosmic-rays of highest energy, possibly $10^{15}$ eV protons. In contrast, the lack of direct association between particle acceleration and field amplification in the dynamo mechanism here does not disfavor it as an interpretation of the $X$-ray rims in SNRs: the particle acceleration might proceed through the motional electric field $E = |{\bf v} \times {\bf B}|/c$ along the rippled shock surface (such an acceleration is known to be faster than in the parallel field case \citep{j87}). We note that the scale of the smallest eddies, or highest turbulent field, inferred here ($\ell_F \sim 10^{16}$ cm) is comparable to the gyroscale $r_g$ of energetic protons with energy $E = 10^{15}$ eV in a field $B = 300 \mu$G ($r_g = 10^{16}$ cm); hence, the mean free path $\lambda$ of the highest energy particles, except the Bohm scattering case, i.e., $\lambda \sim r_g$, is typically greater than $\ell_F$. If this happens, the turbulence downstream is able to enhance scattering of highest energy particles, thereby leading to an additional particle acceleration at the shock possibly beyond the knee of the cosmic-ray spectrum.

\section{Conclusion} \label{sect_concl}

We have investigated the generation of vorticity downstream of a non-relativistic two-dimensional rippled shock front typical of shell-like supernova remnant expanding in a turbulent interstellar medium. By using only jump condition at a rippled shock surface, we have derived the temporal evolution and the saturation of the turbulent magnetic field downstream of the shock, including the non-linear field back-reaction. We conclude that the saturation of $B$ by small-scale dynamo action depends on the shock speed, on the thickness of the layer of the ISM clumps and on the shock curvature radius. The saturation value is found counter-intuitively not to depend on the size of the ISM clumps, as also found for purely hydrodynamic shocks \citep{kmc94}. Perpendicular and parallel field upstream cases lead to the same saturation values, if amplification is very strong, due to fast isotropization of the downstream turbulence. 

Our finding shows that small-scale dynamo might explain non-thermal $X$-ray and optical observations of young/middle-aged supernova remnant shocks. The secular evolution of the turbulent magnetic field derived here might help to shed light on the evolution of young SNRs to be discovered. The youngest SNRs in our galaxy or other ``historical'' SNRs might be in the growing phase of magnetic field described here. Current and next generation of hard X-rays observatories (NuSTAR, Astro-H) will provide a helpful probe for the mechanism here described.
 
%{\it Acknowledgment -- } 

\acknowledgments

The author thanks J. Giacalone and J. R. Jokipii for helpful discussions, 
P. Prasad and N. Kevlahan for correspondence on moving surface theories, 
R. Kulsrud, J. Raymond for comments, Observatory Paris-Meudon, where part of this work was done, 
and the anonymous referee for helpful comments which significantly improved the manuscript.
The support from NASA through the Grants NNX10AF24G and NNX11AO64G is gratefully acknowledged.

\appendix

\section{Computation of the vorticity}

The $\omega_z$ generated downstream of the shock is found, for convenience, by computing separately the jump across the shock of every term on the right side in Eq.(\ref{omegans}), with the shock profile being frozen. Recalling that if ${\bf B } \neq 0$ 
the transverse flow velocity is not conserved across the shock $(\delta v_s \neq 0)$, we find:
\begin{equation}
\delta \omega_z = \partial_s (\delta v_n) - \delta(\partial_n v_s) + (\delta v_s) \partial_s \vartheta  .
\label{omegans2} 
\end{equation}

We compute the term $\delta(\partial_n v_s)$ by using the equations of motion, i.e. Eqs.(\ref{Euler}, \ref{induction}), and following the algorithm used in \citet{k97}. The magnetic field ${\bf B } = (B_n, B_s, 0) \neq 0$ accounts for the back-reaction to the vorticity growth (see Eq.(\ref{dom_perp})). Rewriting the two components of Eq.(\ref{Euler}) in coordinates $(n,s)$, one finds two new equations:
\begin{equation}
\left\{  \begin{array}{cc}
    {\rm cos} \vartheta~ {\mathcal F} + {\rm sin} \vartheta ~{\mathcal G} = 0 \\
    {\rm sin} \vartheta ~{\mathcal F} - {\rm cos} \vartheta ~{\mathcal G} = 0
    \label{eq_theta}
   \end{array}
\right.
\end{equation}
where we have defined
\begin{eqnarray}
{\mathcal F} & = & \frac{d v_n}{dt} + v_s  \frac{d\vartheta}{dt} +  \frac{\partial_n P}{\rho} +   \frac{B_s}{4\pi\rho} (\partial_n B_s - \partial_s B_n  - B_s \partial_s \vartheta)   \nonumber\\  
{\mathcal G} & = & \frac{d v_s}{dt} - v_n  \frac{d\vartheta}{dt} +  \frac{\partial_s P}{\rho} -   \frac{B_n}{4\pi\rho} (\partial_n B_s - \partial_s B_n  - B_s \partial_s \vartheta),  \nonumber
\end{eqnarray}
and $d/dt = \partial_t +v_n \partial_n +v_s \partial_s $.

Equation (\ref{eq_theta}) implies ${\mathcal F} = {\mathcal G} =0$. We recast ${\mathcal G} = 0$ as (see also Sect. 2 in \citet{rp93})
\begin{eqnarray}
\frac{d \vartheta}{dt} & - & \frac{1}{v_n} \frac{d v_s}{dt}  - \frac{\partial_s P}{\rho v_n} 
 \nonumber\\
& - & \frac{B_n}{4\pi\rho v_n} \left(-\partial_n B_s + \partial_s B_n + B_s \partial_s \vartheta \right)=0 .
\label{theta}
\end{eqnarray}
We now evaluate the jump of Eq.(\ref{theta}) across the shock ($\delta (\partial_s \vartheta) = 0$). Multiplying Eq.(\ref{theta}) by $(\rho v_n)_d$ and making use of the mass conservation ($\delta [\rho(v_n - C)] = -\delta [\rho C_r] = 0$), 
we obtain
\begin{align}
& (\rho v_n)_d \delta (\partial_n v_s) + \delta (\rho \partial_t v_s + \rho v_s \partial_s v_s) - (\rho v_n)_d (\delta v_s) \partial_s \vartheta  \nonumber\\
& + \partial_s \delta P + C \delta\rho (\partial_n v_s)_u - C\delta \rho(\partial_t \vartheta)_u - C\delta \rho (v_s)_u \partial_s \vartheta  \nonumber\\ 
& + \frac{B_n}{4\pi} \delta[-\partial_n B_s + \partial_s B_n + B_s \partial_s \vartheta] = 0  \, .
\label{vnd}
\end{align}
Note that if ${\bf B} = 0$ this expression is equivalent to Eq.s [2.8-2.10] in \citet{k97} with $\partial_n \vartheta = 0$.

In Eq.(\ref{vnd}) the mass conservation readily gives: $(\rho v_n)_d (\delta v_s) \partial_s \vartheta + C\delta \rho (v_s)_u \partial_s \vartheta = \delta(\rho v_n v_s) \partial_s \vartheta$. We make here the assumption that the shock velocity in the tangential direction is much smaller than in the normal direction, i.e., obliqueness effects are neglected; hence the jump in transverse momentum writes $\delta[\rho(v_n - C)v_s - B_n B_s/4\pi] = 0$, implying $\delta(\rho v_n v_s) = C\delta(\rho v_s) + B_n \delta B_s/4\pi $. 
Differentiating with respect to $s$ the jump condition of the normal momentum, i.e., $\delta[\rho(C_r)^2 + P - B^2/8\pi] = 0$, provides
\begin{equation}
-\partial_s \delta P = -\partial_s [\rho C_r \delta v_n]  + \partial_s \delta[B^2/8\pi] .
\label{deltaP}
\end{equation}

We replace the expression for $\partial_s \delta P$ from Eq.(\ref{deltaP}) into Eq.(\ref{vnd}) and substitute the resulting expression (\ref{vnd}) for $\delta (\partial_n v_s) $ into the vorticity downstream (Eq. (\ref{omegans2})). By using the jump condition $\delta v_n = \frac{r-1}{r}[C_r]_u$, one can write
\begin{align}
\delta \omega_z & = -\frac{1}{\rho C_r} \left\{ \delta\left[\rho \hat v_s \cdot \left( \frac{D {\bf v}}{Dt }\right) \right]  + \frac{r-1}{r} [C_r]_u \partial_s (\rho C_r)   \right.\nonumber\\
& \quad \left. -  \frac{\partial_s \delta B^2}{ 8\pi}  +  2  \frac{B_n \delta B_s}{4\pi}\partial_s \vartheta + \frac{B_n}{4\pi} \delta[\partial_s B_n - \partial_n B_s ]  \right\} ,
\label{deltaomega}
\end{align}
where $ (D {\bf v}/Dt ) = (\partial_t + v_s  \partial_s + C  \partial_n) {\bf v} $. Note that, if ${\bf B} = 0$, the second line in Eq.(\ref{deltaomega}) vanishes and the remaining terms correspond to the result in \citet{k97}. 

In Eq.(\ref{deltaomega}) we can approximate to first order in $\vartheta$ the shear term: $\hat v_s \cdot (D {\bf v}/Dt ) \sim C_r (D \vartheta/Dt )$. The mass conservation $\delta (\rho C_r) = 0$ for the first term in Eq.(\ref{deltaomega}) gives $\delta[\rho \hat v_s \cdot ( D {\bf v}/Dt ) ] \sim \rho C_r \delta [D\vartheta/Dt] \sim \rho C_r \delta(v_s) \partial_s \vartheta$. The time-derivative $d\vartheta/dt \sim D\vartheta/Dt$ for an MHD shock wave depends on shock speed, magnetic field components and density gradient in a complicated manner \citep{p01}. Remarkably, in the shear term of Eq.(\ref{deltaomega}) the only non-vanishing jump is $(\delta v_s) \partial_s \vartheta$ thus independent on the particular equation of motion for $\vartheta$.

Using the MHD Rankine-Hugoniot conditions for the transverse velocity, $\delta v_s = -B_n \delta B_s/(4\pi \rho C_r)$, we can recast Eq.(\ref{deltaomega}) as Eq.(\ref{deltaomega2}) in the main text.

\end{document}